\documentclass[preprint,tightenlines,aps,prd,showpacs,superscriptaddress,nofootinbib,floatfix]{revtex4}
\usepackage{amsfonts}
\usepackage{multirow}
\usepackage{mathrsfs}
\usepackage{graphicx}
\usepackage{amsmath}
\usepackage{amssymb}
\usepackage{bm}
\usepackage{bbm}
\usepackage{slashbox}
\usepackage{cancel}

\renewcommand{\l}{\left}
\renewcommand{\r}{\right}

\newcommand{\tn}{\textnormal}

\newcommand{\av}[1]{\l|\mathbf{#1}\r|}
\newcommand{\avb}[1]{\l|\boldsymbol{#1}\r|}

\begin{document}
\title{Finite volume corrections to the binding energy of
the $X(3872)$}

\author{M. Jansen}
\affiliation{Institut f\"ur Kernphysik,
Technische Universit\"at Darmstadt, 64289 Darmstadt, Germany}

\author{H.-W. Hammer}
\affiliation{Institut f\"ur Kernphysik,
Technische Universit\"at Darmstadt, 64289 Darmstadt, Germany}
\affiliation{ExtreMe Matter Institute EMMI, GSI Helmholtzzentrum
f\"ur Schwerionenforschung GmbH, 64291 Darmstadt, Germany}

\author{Yu Jia}
\affiliation{Institute of High Energy Physics and Theoretical Physics Center for
Science Facilities, Chinese Academy of
Sciences, Beijing 100049, China}
\affiliation{Center
for High Energy Physics, Peking University, Beijing 100871, China}

\date{\today}
\begin{abstract}
The quark mass dependence of hadrons is an important input for
lattice calculations. We investigate the light quark mass dependence of the
binding energy of the $X(3872)$ in a finite box
to next-to-leading order in an effective field theory for
the $X(3872)$ with perturbative pions (XEFT).
At this order, the quark mass dependence
is determined by a quark mass-dependent contact interaction in addition
to the one-pion exchange. While there is only a moderate sensitivity to
the light quark masses in the region up to twice their physical value, the
finite volume effects are significant already at box length as large as
20~fm.
\end{abstract}



\maketitle
\section{Introduction}
The discovery of the $X(3872)$ in 2003 by the Belle Collaboration \cite{Choi:2003ue} with its confirmation
by the CDF collaboration shortly after \cite{Acosta:2003zx}, was the first of a series of discoveries of
charmonium-like hadrons \cite{Brambilla:2010cs}. Its decays into $J/\psi\pi^+\pi^-$ \cite{Choi:2011fc} and
$J/\psi\pi^+\pi^-\pi^0$ \cite{delAmoSanchez:2010jr} with a ratio of branching fractions close to one,
indicate a large isospin violation in $X$ decays and make an interpretation as a conventional $c\bar{c}$ state
implausible. The description as a loosely bound $\bar{D}^0D^{*0}$ $S$-wave hadronic
molecule with even charge parity \cite{Close:2003sg,Pakvasa:2003ea,Voloshin:2003nt,Wong:2003xk,Braaten:2003he,Swanson:2003tb},
on the other hand, naturally explains the proximity of the $X$ mass to the $\bar{D}^0D^{*0}$ threshold and the
quantum numbers $J^{PC}=1^{++}$ \cite{Aubert:2006aj,Aaij:2013zoa}.

In the molecular picture, the binding energy, $E_X$, is determined by the masses of the $X(3872)$,
the $D^0$ and the $D^{*0}$ meson, $M_X$, $m_D$ and $m_{D^*}$, respectively. Using the
latest values from the review of particle properties \cite{Agashe:2014kda}, it reads
\begin{equation}
	E_X=m_{D^*}+m_{D}-M_X=(0.11\pm 0.21)\tn{ MeV}.
	\label{eq:mxmdmdst}
\end{equation}
A natural energy scale is set by the one-pion exchange, $m_\pi^2/(2M_{DD^*})\approx 10$ MeV,
where $M_{DD^*}$ is the reduced mass of the $D^0$ and $D^{*0}$ mesons and $m_\pi$ the neutral pion mass. The binding energy is small
compared to this natural energy scale and hence, the $X(3872)$ displays universal properties.

Braaten and Kusunoki exploited this universality in a series of papers on the $X$ using effective
field theory methods \cite{Braaten:2003he}. They obtained various predictions for production amplitudes,
decays, formation, and line shapes of the $X(3872)$ (see Ref.~\cite{Braaten:2009zz} for
a review). In Ref.~\cite{AlFiky:2005jd}, the binding energy of the $X(3872)$
was calculated in a pionless effective field theory using constraints from heavy-quark
symmetry.
The influence of three-body $D\bar{D}\pi$ interactions on the properties of the $X(3872)$ was found to be
moderate in a Faddeev approach \cite{Baru:2011rs}.
Finally, we note that universality also determines the interactions of the $X(3872)$ with neutral $D$ and $D^{*}$ mesons
\cite{Canham:2009zq}.
The corrections to universality can be calculated systematically using an effective field theory for the $X$
with explicit pions, called XEFT, which was developed by Fleming, Kusunoki, Mehen and van Kolck in 2007
\cite{Fleming:2007rp}. They applied XEFT to calculate the partial decay width
$\Gamma\l[X\rightarrow D^0\bar{D}^0\pi^0\r]$ at next-to-leading
order (NLO) in the XEFT power counting. Later, their work was extended to describe hadronic decays of the $X(3872)$
to the $\chi_{cJ}$ \cite{Fleming:2011xa}. In Ref.~\cite{Braaten:2010mg}, it was pointed out that
XEFT can also be extended to systems with an additional pion with energies close to the $D^* \bar{D}^*$ threshold.
Finally, a Galilean-invariant formulation of XEFT was introduced in Ref.~\cite{Braaten:2015tga}
to exploit the fact that mass is very nearly conserved in the transition $D^{*0} \to D^0 \pi^0$.

Whereas the $X(3872)$ is appealing for effective field theory approaches particularly because of its
unnatural size, its large extent poses a severe technical problem for lattice simulations.
However, a quenched lattice calculation supported the $J^{PC}=1^{++}$ hypothesis \cite{Yang:2012mya}
before the LHCb experiment finally settled the $X$ quantum numbers \cite{Aaij:2013zoa}. A full lattice QCD
study was first performed by Prelovsek and Lescovec in 2013 \cite{Prelovsek:2013cra}. In this calculation
a candidate for the $X(3872)$ about $(11\pm7)\tn{ MeV}$ below the $\bar{D}^0D^{*0}$ threshold has been found.
The authors used light quark masses at about four times the physical value and a spatial lattice size of only
$L=2\tn{ fm}$. The typical length scale of the $X$ can be estimated from the $\bar{D}^0 D^{*0}$ $S$-wave scattering
length at leading order: $a_s= 1/\sqrt{2M_{DD^*}E_X}$ \cite{Braaten:2004rn}. For the $X$,
this implies $a_s\gtrsim5\tn{ fm}\gg L$, such that large chiral and finite volume effects are expected for the
calculation of \cite{Prelovsek:2013cra}.
Two recent lattice studies \cite{Padmanath:2015era,Lee:2014uta} use similar volumes and pion masses such that
these problems persist.
While the quark mass dependence has been addressed in Refs.~\cite{Wang:2013kva,Baru:2013rta,Jansen:2013cba,Baru:2015nea},
a calculation of the finite volume corrections to observables of the $X(3872)$ is still outstanding.

For two particles in a finite volume, L\"uscher has developed a framework to determine bound-state and scattering observables
from finite volume energy levels
\cite{Luscher:1985dn,Luscher:1986pf}. An equivalent pionless effective field theory approach for two nucleons on a lattice
using the power divergence subtraction (PDS) scheme was presented in \cite{Beane:2003da}. In the quark mass range where the $D^{*0}$ is unstable,
however, the $\bar{D}^0D^{*0}$ system has on-shell $\bar{D}^0D^0\pi^0$ intermediate states and significant three-body effects
are expected.
In the last years, several attempts have been made to obtain a better understanding of the three-body system in a box.
The modification of three-body bound states in a cubic volume was investigated in pionless effective field theory
\cite{Kreuzer:2008bi,Kreuzer:2010ti}.
In Refs. \cite{Polejaeva:2012ut,Briceno:2012rv} and \cite{Hansen:2014eka}, it was shown that finite volume observables are
determined by the infinite volume $S$-matrix elements as it is the case for the two-body system in a box.
There are also some explicit calculations available for systems possessing three-body intermediate states. The $a_1(1260)$
resonance in a cubic box, generated by $\pi\rho$ scattering and accounting for the $\rho$ meson's self energy, has been
addressed in \cite{Roca:2012rx}. Further, in Ref.~\cite{Bour:2011ef}, topological effects for bound states in a moving frame
have been considered. An analytical expression for the finite volume energy shift of three identical bosons in the unitary limit
was derived in Ref.~\cite{Meissner:2014dea}.

In this work, we investigate finite volume corrections to the binding energy of the $X(3872)$. We present explicit expressions
to extrapolate results for the binding energy of the $X(3872)$ from the finite to the infinite volume and from unphysical to
physical quark masses. We consider both, quark masses for which the $D^{*0}$ is stable and quark masses for which it can decay
into $D^0\pi^0$. Furthermore, we use the quark mass dependence synonymous to the pion mass dependence because of the Gell-Mann-Oakes-Renner
relation \cite{GellMann:1968rz}:
\begin{equation}
m_\pi^2 = -(m_u + m_d) \langle 0 | \bar{u} u + \bar{d}d| 0 \rangle /f^2\,,
\label{eq:mq-mpi}
\end{equation}
where $f\approx 130$ MeV is the pion decay constant,
$m_u$ and  $m_d$ are the light quark masses, and
$\langle 0 | \bar{u} u | 0 \rangle = \langle 0 | \bar{d} d | 0 \rangle = (-283(2) \mbox{ MeV})^3$ is the light quark condensate
in the $\overline{MS}$ scheme at 2 GeV \cite{McNeile:2012xh}.

The paper is organized as follows: In Sec.~\ref{sec:infvol}, we briefly review XEFT in the infinite volume. A strategy how to obtain shifts
for the binding energy due to higher order contributions is presented in Sec.~\ref{sec:BNLO}. The finite volume amplitudes are calculated
in Sec.~\ref{sec:finvol} and explicit expressions for the binding energy in dependence on the box size are given. In Sec.\ref{sec:results},
we discuss our results for the chiral and finite volume extrapolations of the $X(3872)$ binding energy.
Finally, we summarize our conclusions and present an outlook on future work in Sec.~\ref{sec:conout}.

\section{Review of XEFT in the infinite volume}
\label{sec:infvol}
We briefly review the results for the binding energy in the infinite volume, following our analysis in \cite{Jansen:2013cba}. The underlying
effective theory we are using in order to describe the $X$, called XEFT, was derived by Fleming et. al. in \cite{Fleming:2007rp} starting from heavy-meson
chiral perturbation theory. In XEFT, regarding the $X(3872)$ as a $\bar{D}^0D^{*0}$ $S$-wave hadronic molecule, $D^0$, $D^{*0}$, $\bar{D}^0$, $\bar{D}^{*0}$
and $\pi^0$ fields are treated non-relativistically. Charged $D^{(*)}$ mesons are integrated out and do not contribute to the order we are working at.
Moreover, the $J/\psi\rho$ and $J/\psi\omega$ channels can also be integrated out~\cite{Fleming:2007rp}.
Utilizing XEFT, we can evaluate the center-of-momentum $\bar{D}^0D^{*0}$ $S$-wave scattering diagrams and eventually extract the binding energy of the $X$.

The XEFT Lagrangian reads
\begin{align}
  \mathcal{L}=&\boldsymbol{D}^\dagger\l(i\partial_0+\frac{\overrightarrow{\nabla}^2}{2m_{D^*}}\r)
  \boldsymbol{D}+D^\dagger\l(i\partial_0+\frac{\overrightarrow{\nabla}^2}{2m_D}\r)D\notag\\
  +&\boldsymbol{\bar{D}}^\dagger\l(i\partial_0+\frac{\overrightarrow{\nabla}^2}{2m_{D^*}}\r)
  \boldsymbol{\bar{D}}+\bar{D}^\dagger\l(i\partial_0+\frac{\overrightarrow{\nabla}^2}{2m_D}\r)\bar{D}+\pi^\dagger\l(
  i\partial_0+\frac{\overrightarrow{\nabla}^2}{2m_\pi}+\delta\r)\pi\notag\\
  +&\frac{g}{\sqrt{2}f}\frac{1}{\sqrt{2m_\pi}}\l(D\boldsymbol{D}^\dagger\cdot\overrightarrow{\nabla}\pi+\bar{D}^\dagger
  \boldsymbol{\bar{D}}\cdot\overrightarrow{\nabla}\pi^\dagger\r)+\tn{h.c.}\notag\\
  -&\frac{C_0}{2}\l(\boldsymbol{\bar{D}}D+\boldsymbol{D}\bar{D}\r)^\dagger\cdot\l(\boldsymbol{\bar{D}}D+
  \boldsymbol{D}\bar{D}\r)\notag\\
  +&\frac{C_2}{16}\l(\boldsymbol{\bar{D}}D+\boldsymbol{D}\bar{D}\r)^\dagger\cdot\l(
  \boldsymbol{\bar{D}}\overleftrightarrow{\nabla}^2D+\boldsymbol{D}\overleftrightarrow{\nabla}^2\bar{D}\r)+\tn{h.c.}\notag\\
  -&\frac{D_2\mu^2}{2}\l(\boldsymbol{\bar{D}}D+\boldsymbol{D}\bar{D}\r)^\dagger\cdot\l(\boldsymbol{\bar{D}}D+
  \boldsymbol{D}\bar{D}\r)+\cdots.
  \label{eq:xeftlagrangian}
\end{align}
Here, $\overleftrightarrow{\nabla}\equiv(\overrightarrow{m}\overleftarrow{\nabla}-\overleftarrow{m}\overrightarrow{\nabla})/(\overleftarrow{m}+\overrightarrow{m})$ is the Galilean invariant derivative, $\overleftarrow{m}$ ($\overrightarrow{m}$) the mass of the left- (right-) hand field
and the ellipsis denote higher-order interactions. The masses for the pion, the $D^0$- and $D^{*0}$-meson are labeled by $m_\pi$, $m_D$ and $m_{D^*}$, respectively. Furthermore, $g$ is the $D$ meson axial coupling constant, $f$ the pion decay constant and $\Delta\equiv m_{D^*}-m_D$ the hyperfine splitting of the $D$ mesons. The mass scales $\mu$ and $\delta$ are defined as $\mu^2\equiv\Delta^2-m_\pi^2$ and $\delta\equiv\Delta-m_\pi$. The coupling constants $C_0$, $C_2$ and $D_2$ are discussed below.
Note that this Lagrangian has no exact Galilean invariance. The $D^{*0}D^0\pi^0$ coupling is given by the leading term in the chiral limit and only the
leading terms in expansion in $m_\pi/m_D$ and $\delta/m_\pi$ are kept in the calculated observables.

In its structure, XEFT is similar to the Kaplan-Savage-Wise theory (KSW) for nucleon-nucleon ($NN$) scattering, which makes use of the power divergence subtraction scheme (PDS) \cite{Kaplan:1998tg}. PDS has proven to be well suited for systems with an unnaturally large scattering length.
In the KSW counting, the pion exchanges are included perturbatively. Although the perturbative treatment of the pion exchanges was shown to fail in the $NN$ sector
at NNLO because of large contributions from the nuclear tensor force \cite{Fleming:1999ee}, it is expected to work well for $\bar{D}^0D^{*0}$ scattering in XEFT due to a significantly smaller expansion parameter \cite{Fleming:2007rp}. In XEFT, the mass scale $\mu$, the momenta of the $D^{0(*)}$ mesons and the pions as well as the binding momentum are all counted as order $Q$, which defines the typical momentum scale of XEFT. At leading order (LO), that is $Q^{-1}$ in XEFT power counting, there is one contact interaction with coupling constant $C_0$. Loop integrations contribute a factor $Q^5$ and propagators a factor $Q^{-2}$. Therefore a loop ($Q^5$) consisting of two propagators (each $Q^{-2}$) and a LO contact interaction ($Q^{-1}$) is counted as order $Q^0$. Appending such a loop to any diagram leaves the order of the diagram unchanged.
On the one hand, this implies that the LO contact interaction has to be resummed to all orders. On the other hand, the higher order contributions
have to be dressed in all possible ways by LO amplitudes. The NLO amplitude is of order $Q^0$ and three more interactions have to be considered: two NLO contact interactions with coupling constants $C_2$ and $D_2$ and the $D^{*0}D^0\pi^0$ coupling. 

Further, XEFT explicitly accounts for the finite decay width of the $D^{*0}$ meson. Whereas nucleons, regarding strong interactions only, are stable, the $D^{*0}$ can decay into $D^0\pi$ as long as the hyperfine splitting of the $D^{*0}$ and the $D^0$ meson is greater than the pion mass.
Thus additional diagrams have to be included which lead to the emergence of infrared (IR) divergences. In order to cure this pathological behavior, we resum pion contributions to the propagator of the $D^{*0}$ meson \cite{Jansen:2013cba}. Inserting these resummed instead of bare $D^{*0}$ propagators in all $\bar{D}^0D^{*0}$ loops, we obtain IR finite amplitudes.
For more details on infrared divergences, the XEFT power counting and the derivation of the Lagrangian, we refer to \cite{Fleming:2007rp} and \cite{Jansen:2013cba}.

In the Galilean-invariant version of XEFT from \cite{Braaten:2015tga}, the width from decays of the $D^{*0}$ into both $D^0\pi^0$ and $D^0\gamma$ is included
and the Galilean invariance is exact. It strongly constrains the form of the ultraviolet divergences in the theory such that no expansion in $m_\pi/m_D$
is required. A covariant formulation of a non-relativistic effective field theory describing Goldstone boson dynamics can be found in
Ref.~\cite{Colangelo:2006va}.

\begin{figure}[htbp]
	\begin{center}
		\includegraphics[width=0.95\textwidth]{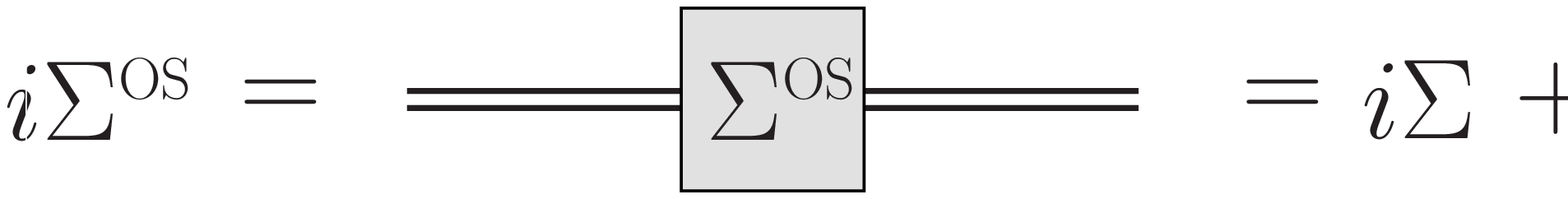}
	\end{center}
	\caption{Self-energy graph with pion contribution and counterterm for the $D^{*0}$. Single, double and dashed lines represent $D^0$, 
$D^{*0}$ and pion propagators, respectively. The cross on the right hand side indicates the insertion of the counterterm.}
	\label{fig:isigma}
\end{figure}

\begin{figure}[htbp]
	\begin{center}
		\includegraphics[width=0.8\textwidth]{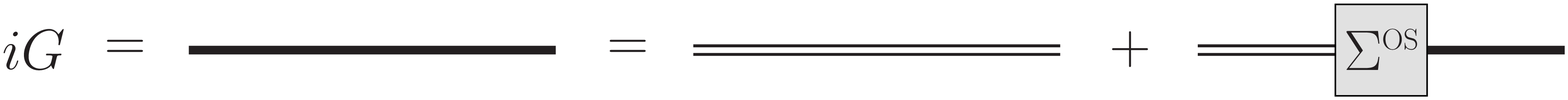}
	\end{center}
	\caption{Fully resummed $D^{*0}$ propagator with the insertion of the self-energy $i\Sigma^\tn{OS}$. Bare and full propagator are represented by double and thick lines, respectively.}
	\label{fig:Zds2}
\end{figure}

We begin with the derivation of the resummed $D^{*0}$ propagator, utilized in all calculations for the $\bar{D}^0D^{*0}$ system.
First, we consider the $D^{*0}$ self-energy shown in Fig. \ref{fig:isigma}. For the counterterm, $i\delta_\Sigma$, we use the on-shell renormalization scheme. It ensures that the real part of the propagator's pole position is at the on-shell point $p_0=p^2/2m_{D^*}$, with $p_0$ being the energy and $p$ the momentum of the $D^{*0}$. 
Using PDS\footnote{
Note that the additional term occurring in PDS, proportional to the renormalization scale $\Lambda$, is subtracted again due to the use of on-shell renormalization scheme.}, the bare self-energy reads
\begin{align}
	i\Sigma=\frac{ig^2}{24\pi f^2}\l(i\mu^3+\Lambda\mu^2\r),
	\label{eq:bareselfenergy}
\end{align}
where $\Lambda$ is the PDS renormalization scale. The counterterm is chosen such that it cancels the second term in parentheses, which is real valued, analytic in the quark masses and proportional to the PDS renormalization scale $\Lambda$. The first term is purely imaginary as long as the $D^{*0}$ can decay into $D^0\pi$, i.e. for pion masses smaller than the hyperfine splitting. In this case it induces a finite decay width. For pion masses greater than the hyperfine splitting, it is real valued and the self-energy implies a finite mass shift for the $D^{*0}$, denoted by $\Delta_{m_{D^*}}$.
In summary we have
\begin{align}
	i\delta_\Sigma&=-\frac{ig^2}{24\pi f^2}\Lambda\mu^2,\\
	\Delta_{m_{D^*}}&=\begin{cases}
		0,&m_\pi<\Delta,\\
		\frac{g^2}{24\pi f^2}i\mu^3,&m_\pi\geq\Delta.
	\end{cases}
	\label{eq:countertermmassshift}
\end{align}
We point out that the mass scale $\mu$ is purely imaginary above the $D^{*0}$ decay threshold and hence $\Delta_{m_{D^*}}$ is real valued for all pion masses. The $D^{*0}$ propagator can now be calculated according to Fig. \ref{fig:Zds2} and we obtain after resummation
\begin{align}
	iG=\frac{i}{p_0-p^2/2m_{D^*}+\Sigma^\tn{OS}+i\epsilon}.
	\label{eq:Dstarresummed}
\end{align}

\begin{figure}[htbp]
	\begin{center}
		\includegraphics[width=0.8\textwidth]{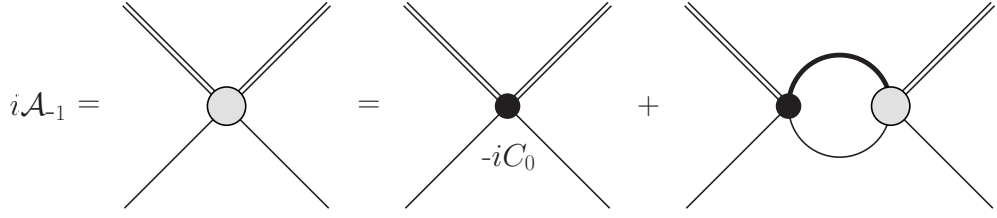}
	\end{center}
	\caption{Leading-order diagram for $\bar{D}^0D^{*0}$ scattering. The single and double lines represent $\bar{D}^0$ and $D^{*0}$ mesons, respectively. The thick, internal line indicates a fully resummed $D^{*0}$ propagator depicted in Fig. \ref{fig:Zds2}.}
	\label{fig:Am1}
\end{figure}

We proceed with the evaluation of the $\bar{D}^0D^{*0}$ scattering diagrams. Since we consider $S$-wave scattering, the total angular momentum of the $\bar{D}^0D^{*0}$ system is determined by the $D^{*0}$ meson's spin. Denoting the polarization vectors of the incoming and outgoing $D^{*0}$ by $\boldsymbol{\varepsilon}_i$ 
and $\boldsymbol{\varepsilon}_j^*$, respectively, it turns out that all amplitudes factorize and we can write
\begin{align}
	\mathcal{A}_{ij}=\delta_{ij}\mathcal{A},
	\label{eq:saseparation}
\end{align}
using spin indices $i$ and $j$. For the calculation of the binding energy, it is sufficient to take a look at the scalar amplitudes $\mathcal{A}$. At leading order, there is one contact interaction with coupling constant $C_0$. According to XEFT power counting, it has to be resummed to all orders, as depicted in Fig.~\ref{fig:Am1}. 
Using PDS to renormalize the linear divergence of the loop integral in Fig. \ref{fig:Am1},
\begin{align}
	I_0=\int\frac{d^3\mathbf{k}}{\l(2\pi\r)^3}\frac{1}{\av{k}^2+\eta^2}\xrightarrow{\tn{PDS}}\frac{1}{4\pi}\l(\Lambda-\eta\r),
	\label{eq:I0inf}
\end{align}
with the energy-dependent quantity $\eta\equiv\sqrt{-2M_{DD^*}(E+\Sigma^\tn{OS})-i\epsilon}$, we obtain
\begin{align}
	i\mathcal{A}_{-1}=\frac{2\pi i}{M_{DD^*}}\frac{1}{-\frac{2\pi}{M_{DD^*}C_0}-4\pi I_0(\eta^2)}=\frac{2\pi i}{M_{DD^*}}\frac{1}{-\gamma+\eta}.
	\label{eq:Am1}
\end{align}
The LO coupling constant occurs in the definition of $\gamma$
\begin{align}
	\gamma\equiv\frac{2\pi}{M_{DD^*}C_0(\Lambda)}+\Lambda.
	\label{eq:gamma}
\end{align}
The LO amplitude has a pole at $-E=B^\tn{LO}\equiv\gamma^2/2M_{DD^*}+\Sigma^\tn{OS}$ and we identify the binding energy with the real part of $B^\tn{LO}$. 
Thus, $\gamma$ is the LO binding momentum, and $C_0$ in Eq. \eqref{eq:gamma} has to cancel the dependence on the 
PDS renormalization scale $\Lambda$ and fix the binding energy for physical quark masses to the experimentally measured value.

At NLO, there are three more interactions, which are included perturbatively. Two contact interactions, with coupling constants $C_2$ and $D_2$, and the one-pion exchange (OPE). Following XEFT power counting, the LO amplitude has to be appended in all possible ways to the NLO interactions.
We end up with the scattering diagrams shown in Fig. \ref{fig:A0}. The results for the $S$-wave projected scalar amplitudes are given in
Appendix \ref{sec:NLOsa}.
\begin{figure}[htbp]
	\begin{center}
		\includegraphics[width=0.8\textwidth]{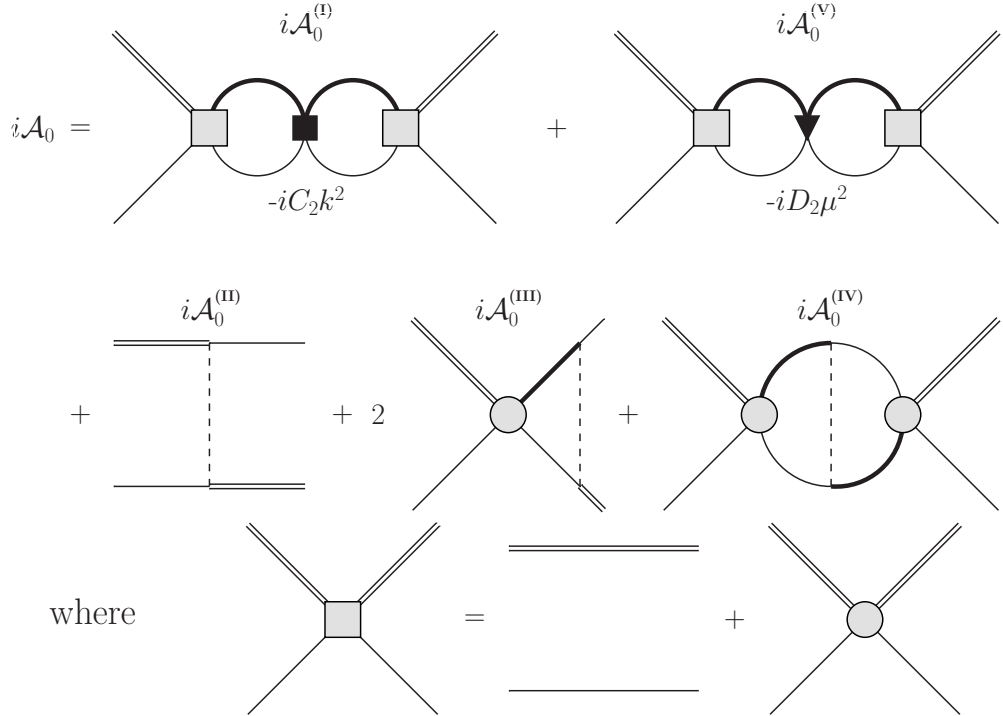}
	\end{center}
	\caption{Next-to-leading-order diagrams for $\bar{D}^0D^{*0}$ scattering. We use the same notation as in Figs. \ref{fig:isigma}, \ref{fig:Zds2} and \ref{fig:Am1}.}
	\label{fig:A0}
\end{figure}

After the inclusion of NLO contributions, the renormalization condition, when fixing the binding energy at its experimentally measured value for physical quark masses, implies a relation between the LO and the NLO coupling constants.

Since $C_2$ and $D_2$ are unknown, we estimate natural ranges and vary the coupling constants to determine the error band of the binding energy. 
We rewrite the coupling constants as
\begin{subequations}
\begin{align}
	C_2&=\frac{M_{DD^*}}{2\pi}\frac{r_0}{2}(C_0)^2\equiv c_2(C_0)^2,\label{eq:C2renor}\\
	D_2&=\frac{g^2}{6f^2}\l(\frac{M_{DD^*}}{2\pi}\r)^2\l(d_2+\log\l(\frac{\Lambda}{\mu^\tn{ph}}\r)+R\r)(C_0)^2,\label{eq:D2renor}
\end{align}
\end{subequations}
where $R$ is a renormalization constant given in Appendix \ref{sec:NLOsa} and the superscript ``ph" indicates that a quantity, here the mass scale $\mu$, is evaluated at 
physical quark masses. Furthermore, compared with the pure contact theories, $r_0$ can be identified with the effective range in the pionless theory. We follow \cite{Fleming:2007rp} and \cite{Jansen:2013cba} and use their estimates $r_0\in[0,1/100\tn{MeV}]$ and $d_2\in[-1,1]$. In the future, it should be possible to determine $C_2$ and $D_2$ from lattice calculations.

\section{Strategy for extracting the binding energy to NLO}
\label{sec:BNLO}

For an unstable $D^{*0}$ meson, the OPE potential is oscillatory and not Yukawa-like \cite{Suzuki:2005ha}. The effective range expansion breaks down
at NLO and the effective range is not defined. Hence, the binding energy can not be extracted from effective range parameters.
In this section, we present an alternative
method to access the binding energy up to NLO, employing the two-body scattering amplitudes, regardless of whether examining stable or unstable particles in the finite or infinite volume. First, we note that the sum of the NLO scattering amplitudes, $\mathcal{A}_0^\tn{(I)},\cdots,\mathcal{A}_0^\tn{(V)}$, can be collected in powers of the LO amplitude
\begin{align}
	\mathcal{A}_0=\mathcal{A}_0^\tn{(I)}+\cdots+\mathcal{A}_0^\tn{(V)}=s_0+s_1\mathcal{A}_{-1}+s_2(\mathcal{A}_{-1})^2.
	\label{eq:collectedA0}
\end{align}
Furthermore, we expand the LO amplitude around the LO pole position
\begin{align}
	\mathcal{A}_{-1}=\frac{Z_{-1}}{E+B^\tn{LO}}+\cdots,
	\label{eq:expandedAm1}
\end{align}
where the dots denote terms being finite at $E=-B^\tn{LO}$ and $Z_{-1}$ is the residue
\begin{align}
	(Z_{-1})^{-1}=\l[i\frac{\partial}{\partial E}\frac{1}{i\mathcal{A}_{-1}}\r]_{E=-B^\tn{LO}}=\frac{-(M_{DD^*})^2}{2\pi}\frac{1}{\gamma}.
	\label{eq:Zm1}
\end{align}
Accordingly, the full amplitude up to NLO, expanded around the LO pole position, can be written as
\begin{align}
	\mathcal{A}=\mathcal{A}_{-1}+\mathcal{A}_0=\frac{Z_{-1}+s_1Z_{-1}}{E+B^\tn{LO}}+\frac{s_2(Z_{-1})^2}{(E+B^\tn{LO})^2}+\cdots.
	\label{eq:fullamplitude}
\end{align}
Moreover, we consider a generic, non-perturbative expression for the amplitude with shifted pole position, $B=B^\tn{LO}+\Delta B$, and shifted residue, $Z=Z_{-1}+\Delta Z$, and expand it around the LO pole position
\begin{align}
	\mathcal{A}^\tn{np}=\frac{Z}{E+B}+\cdots=\frac{Z_{-1}+\Delta Z}{E+B^\tn{LO}}-\frac{Z\Delta B}{(E+B^\tn{LO})^2}+\cdots.
	\label{eq:nonpertamplitude}
\end{align}
Utilizing expressions \eqref{eq:fullamplitude} and \eqref{eq:nonpertamplitude}, the NLO shifts for the residue, $Z_{-1}$, and the LO pole position, $B^\tn{LO}$, can be read off by equating coefficients
\begin{subequations}
\label{eq:BZshift}
	\begin{align}
		\Delta Z^\tn{NLO}&=s_1Z_{-1},\label{eq:Zshift}\\
		\Delta B^\tn{NLO}&=-\frac{s_2}{1+s_1}Z_{-1}\xrightarrow{\cancel{\tn{NNLO}}}-s_2Z_{-1},\label{eq:Bshift}
	\end{align}
\end{subequations}
where we already used a partial NNLO cancellation in Eq.~\eqref{eq:Bshift}, which is described in Appendix \ref{sec:NNLOcancel} in more detail. For a pure contact theory, a comparison to an approach, where the NLO coupling constants are resummed to all orders, can be found in Sec.~\ref{sec:results}. In the following, we apply this
strategy to extract the binding energy of the $X(3872)$ in a finite volume.

\section{Finite volume corrections to the binding energy}
\label{sec:finvol}

We consider the $\bar{D}^0D^{*0}$ system in a box with side length $L$ and periodic boundary conditions. The allowed lattice momenta are then given by integer vectors times $2\pi/L$. Integrals, occurring in calculations for the binding energy in the infinite volume, have to be replaced by discrete sums over the quantized lattice momenta. Since we are interested in $\bar{D}^0D^{*0}$ bound states with even parity, we expect that the binding energy acquires a positive shift. We distinguish between two different regions. One, where the $D^{*0}$ is unstable, i.e. for pion masses $m_\pi<\Delta$ and a second region where the $D^{*0}$ is stable, i.e. $m_\pi\geq\Delta$. Whereas in the first case explicit XEFT calculations have to be carried out due to three-body $D\bar{D}\pi$ intermediate states, in the latter case one can alternatively use a two-body approach introduced in \cite{Beane:2003da}, which serves as a consistency check. All quantities which differ in the finite volume are tagged by a superscript $L$.

\subsection{The LO amplitude}
\label{sec:LOXEFTfinvol}
Let us begin with the explicit XEFT calculations, which can be utilized in both regions. To LO, the amplitude in the finite volume reads
\begin{align}
	i\mathcal{A}_{-1}^L=\frac{2\pi i}{M^L_{DD^*}}\frac{1}{-\frac{2\pi}{M^L_{DD^*}C_0}-4\pi I_0^L},
	\label{eq:Am1L}
\end{align}
where the finite volume quantity $I_0^L$ is given by
\begin{align}
	I_0^L\equiv\frac{1}{L^3}\sum_{\mathbf{k}=\frac{2\pi}{L}\mathbf{n}}\frac{1}{\av{k}^2-2M^L_{DD^*}E},~\mathbf{n}\in\mathbb{Z}^3.
	\label{eq:I0L}
\end{align}
Since the mass of the $D^{*0}$ meson obtains a shift in the finite volume as given below, the reduced mass is dependent on the box size, too.
Like the loop integral \eqref{eq:I0inf} in the infinite volume, $I_0^L$ is linearly ultraviolet divergent since short-distance properties of the theory remain unchanged in the finite volume. We regularize following \cite{Beane:2003da}: first we introduce a sharp momentum cut-off, $\lambda$, for the sum and then add and subtract the infinite volume loop integrals evaluated at zero energy. One of the loop integrals is regularized using PDS, the other one using a momentum cut-off, which coincides with the cut-off in the sum. Finally, the limit $\lambda\rightarrow\infty$ is taken. We obtain
\begin{align}
	I_0^L\xrightarrow{\tn{PDS}}\lim_{\lambda\rightarrow\infty}\l[\frac{1}{L^3}\sum_{\mathbf{k}=\frac{2\pi}{L}\mathbf{n}}^{\av{k}<\lambda}\frac{1}{\av{k}^2-p^2}-\int\frac{d^3\mathbf{k}}{\l(2\pi\r)^3}\frac{\theta\l(\lambda-\av{k}\r)}{\av{k}^2}\r]+\frac{\Lambda}{4\pi},
	\label{eq:I0Lreg}
\end{align}
with $E=p^2/2M^L_{DD^*}$. Plugging \eqref{eq:I0Lreg} into \eqref{eq:Am1L} and using the definition for the LO binding momentum in the infinite volume
(\ref{eq:gamma}), we acquire
\begin{align}
	i\mathcal{A}_{-1}^L=\frac{-2\pi i}{M^L_{DD^*}}\frac{1}{\gamma+\frac{1}{\pi L}S\l(\l(\frac{Lp}{2\pi}\r)^2\r)},
	\label{eq:Am1Lreg}
\end{align}
with
\begin{align}
	S(x)=\lim_{\lambda_n\rightarrow\infty}\l[\sum_{\mathbf{n}}^{\av{n}<\lambda_n}\frac{1}{\av{n}^2-x}-4\pi\lambda_n\r],
	\label{eq:Sx}
\end{align}
where $\lambda_n\equiv L\lambda/( 2\pi)$.
The energy levels of the $\bar{D}^0D^{*0}$ system to LO in the finite volume can be determined from Eq. \eqref{eq:Am1Lreg}. Note that they are fully determined by the infinite volume quantity $\gamma$.\footnote{We expect that even if three particle intermediate states exist that is after the inclusion of NLO contributions and for $m_\pi<\Delta$, finite volume observables are still determined by the infinite volume $S$-matrix as demonstrated in \cite{Polejaeva:2012ut,Briceno:2012rv,Hansen:2014eka}.} Here, we are interested in the solution with negative energy, i.e. the solution which approaches the infinite volume LO binding energy for $L\rightarrow\infty$. We denote the corresponding LO binding momentum by $\gamma^L$, defined by
\begin{align}
	\gamma+\frac{1}{\pi L}S\l(-\l(\tfrac{L\gamma^L}{2\pi}\r)^2\r)=0.
	\label{eq:bssol}
\end{align}

\subsection{The $D^{*0}$ self-energy and mass shift}
\label{sec:dstarselffinvol}
We proceed with the $D^{*0}$ self-energy and mass shift. The calculation is carried out similarly to the one of the LO amplitude. Using PDS and cut-off regularization we obtain for the bare self-energy
\begin{align}
	\Sigma^L=\frac{g^2}{24\pi (f^L)^2}(\mu^L)^2\l(\frac{1}{\pi L}S\l(\l(\tfrac{L\mu^L}{2\pi}\r)^2\r)+\Lambda\r).
	\label{eq:bareselfenergyL}
\end{align}
Independent of the pion mass, we do not receive any imaginary contributions for the bare self-energy in the box. However, since the finite volume itself cuts off low frequency modes, we do not expect the occurrence of any infrared divergences.

Again, we use the on-shell renormalization scheme and subtract the second term proportional to the PDS renormalization scale $\Lambda$. Hence, the counterterms in the finite and infinite volume coincide up to corrections to $f$ and $\mu$. The shift for the $D^{*0}$ mass is different though,
\begin{align}
	\Delta_{m_{D^*}}^L=\frac{g^2}{24\pi (f^L)^2}(\mu^L)^2\frac{1}{\pi L}S\l(\l(\tfrac{L\mu^L}{2\pi}\r)^2\r).
	\label{eq:deltamdstar}
\end{align}
Note that even for physical pion mass, the $D^{*0}$ meson in a box receives a finite mass shift.

\subsection{NLO corrections to the binding energy}
\label{sec:NLOXEFTfinvol}
Now, we implement the corrections due to the NLO amplitudes. In analogy to the infinite volume we find for the NLO contact interactions, i.e. the amplitudes $\mathcal{A}_{0\tn{(I)}}^L$ and $\mathcal{A}_{0\tn{(V)}}^L$
\begin{subequations}
	\begin{align}
		i\mathcal{A}_{0\tn{(I)}}^L&=\frac{-iC_2p^2}{(C_0)^2}(\mathcal{A}_{-1}^L)^2,\\
		i\mathcal{A}_{0\tn{(V)}}^L&=\frac{-iD_2(\mu^L)^2}{(C_0)^2}(\mathcal{A}_{-1}^L)^2.
		\label{eq:NLOcontact}
	\end{align}
\end{subequations}
For the pion exchange diagrams, we do not project onto the $S$-waves. Whereas the infinite volume 
is rotationally invariant, the lattice is only invariant under transformations of the cubic group. In principle, it is possible to decompose quantities transforming according to an irreducible representation of the cubic group into spherical harmonics \cite{Altmann:1965,Kreuzer:2009jp}. However, we keep the sums over integer vectors, since convergence of the partial wave expansion is not certain. The OPE amplitude $\mathcal{A}_{0\tn{(II)}}$ is then given as
\begin{align}
	i\hat{\mathcal{A}}_{0\tn{(II)}ij}^L=\frac{ig^2}{2(f^L)^2}\frac{(\boldsymbol{\varepsilon}_i\cdot(\boldsymbol{\ell}-\boldsymbol{\ell'}))(\boldsymbol{\varepsilon}_j^*\cdot(\boldsymbol{\ell}-\boldsymbol{\ell'}))}{\l|\boldsymbol{\ell}-\boldsymbol{\ell'}\r|^2-(\mu^L)^2},
	\label{eq:A0II}
\end{align}
with the incoming (outgoing) relative momentum $\boldsymbol{\ell}$ ($\boldsymbol{\ell'}$). For $\mathcal{A}_{0\tn{(III)}}$ we find
\begin{align}
	i\hat{\mathcal{A}}_{0\tn{(III)}ij}^L=\mathcal{A}_{-1}^L\frac{M^L_{DD^*}}{2\pi}\frac{ig^2}{2(f^L)^2}\l(\frac{1}{\pi L}S^\tn{(III)}_{ij}\l(\tfrac{L\boldsymbol{\ell}}{2\pi},\l(\tfrac{L\mu^L}{2\pi}\r)^2\r)+\frac{\delta_{ij}}{3}\Lambda\r)+\boldsymbol{\ell}\longleftrightarrow\boldsymbol{\ell'},
	\label{eq:A0III}
\end{align}
where the quantity $S^\tn{(III)}_{ij}$ is defined as
\begin{align}
	S^\tn{(III)}_{ij}\l(\mathbf{m},x\r)\equiv\lim_{\lambda_n\rightarrow\infty}\l[\sum_{\mathbf{n}}^{\av{n}<\lambda_n}\frac{1}{\av{n}^2-\av{m}^2}\frac{\boldsymbol{\varepsilon}_i\cdot(\mathbf{n}+\mathbf{m})~\boldsymbol{\varepsilon}^*_j\cdot(\mathbf{n}+\mathbf{m})}{\l|\mathbf{n}+\mathbf{m}\r|^2-x}-\frac{\delta_{ij}}{3}4\pi\lambda_n\r].
	\label{eq:SIII}
\end{align}
The amplitudes $\hat{\mathcal{A}}_{0\tn{(II)}ij}^L$ and $\hat{\mathcal{A}}_{0\tn{(III)}ij}^L$ imply a coupling between channels with different angular momentum. Considering the $A_1$ representation of the cubic group, the lowest angular momenta coupled are with $l=0,4,6,8,\cdots$. On the other hand, for the amplitude $\hat{\mathcal{A}}_{0\tn{(IV)ij}}^L$ we can use a tensor decomposition and it appears that $\hat{\mathcal{A}}_{0\tn{(IV)}ij}^L=\delta_{ij}\mathcal{A}_{0\tn{(IV)}}^L$. A detailed derivation is given in Appendix \ref{sec:A0IV}. We obtain for the scalar amplitude
\begin{align}
	i\mathcal{A}_{0\tn{(IV)}}^L=&(\mathcal{A}_{-1}^L)^2\l(\frac{M^L_{DD^*}}{2\pi}\r)^2\frac{ig^2}{6(f^L)^2}\biggl[\l(\frac{1}{\pi L}S\l(\l(\tfrac{Lp}{2\pi}\r)^2\r)+\Lambda\r)^2\notag\\
	+&(\mu^L)^2\biggl(\frac{1}{(2\pi^2)^2}S_\tn{(IV)}\l(\l(\tfrac{Lp}{2\pi}\r)^2,\l(\tfrac{L\mu^L}{2\pi}\r)^2\r)+\log\l(\frac{\Lambda}{\l|\mu^L\r|}\r)+\frac{1}{2}+R\biggr)\biggr],
	\label{eq:ALIV}
\end{align}
where
\begin{align}
	S^\tn{(IV)}(x,y)\equiv\lim_{\lambda_n\rightarrow\infty}\Biggl[&\sum_{\mathbf{n},\mathbf{n}'}^{\av{n},\av{n'}<\lambda_n}\frac{1}{\av{n}^2-x}~\frac{1}{\av{n'}^2-x}~\frac{1}{\l|\mathbf{n}+\mathbf{n'}\r|^2-y}-2\pi^4\l(\log\l(\frac{\lambda_n^2}{\l|y\r|}\r)-1\r)\Biggr].
	\label{eq:SIV2}
\end{align}
Due to the coupling between different angular momenta, the amplitudes $\hat{\mathcal{A}}_{0\tn{(II)}ij}^L$ and $\hat{\mathcal{A}}_{0\tn{(III)}ij}^L$ in the finite volume do not factorize into a scalar amplitude and a function of the incoming and outgoing $D^{*0}$ mesons' spins, in particular $\hat{\mathcal{A}}_{0\tn{(II),(III)}ij}^L\neq\delta_{ij}\mathcal{A}_{0\tn{(II),(III)}}^L$. This implies a non-trivial dependence of the coefficients $s_0$ and $s_1$ in Eq. \eqref{eq:collectedA0} on the polarization vectors $\boldsymbol{\varepsilon_i}$ and $\boldsymbol{\varepsilon_j}$ and hence of the NLO shift for the field strength renormalization constant, $\Delta Z^\tn{NLO}$. However, since the amplitudes $\hat{\mathcal{A}}_{0\tn{(I)}ij}^L$, $\hat{\mathcal{A}}_{0\tn{(IV)}ij}^L$ and $\hat{\mathcal{A}}_{0\tn{(V)}ij}^L$ do factorize\footnote{The factorization takes place since the amplitudes $\hat{\mathcal{A}}_{0\tn{(I)}ij}^L$, $\hat{\mathcal{A}}_{0\tn{(IV)}ij}^L$ and $\hat{\mathcal{A}}_{0\tn{(V)}ij}^L$ contain the momentum and hence angular independent LO amplitude $\mathcal{A}_{-1}$ on both sides and thus are momentum and angular independent by themselves.} and therefore $s_2$, it is sufficient to consider the scalar amplitudes $\mathcal{A}_{0\tn{(I)}}^L$, $\mathcal{A}_{0\tn{(IV)}}^L$ and $\mathcal{A}_{0\tn{(V)}}^L$ to calculate the shift for the binding energy.

The dependence of the loop integrals on the PDS renormalization scale $\Lambda$ is the same as for the infinite volume and accordingly the NLO coupling constants coincide with the ones given in Eqs. \eqref{eq:C2renor} and \eqref{eq:D2renor} up to finite volume corrections to scales large compared to $Q$ like for example $m_{D^*}$ or $f$.
This corresponds to a multiplicative renormalization scheme where loop integrals are regularized separately.
Again, the error bands are obtained by varying the coupling constants within their natural ranges.
For the binding energy we employ the results of the previous section. The quantities $Z_{-1}$ and $s_2$ have to be reevaluated in the box. We find for the residue
\begin{align}
	(Z_{-1}^L)^{-1}&=\l[i\frac{\partial}{\partial E}\frac{1}{i\mathcal{A}_{-1}^L}\r]_{E=-B_\tn{LO}^L}=\frac{-\l(\tfrac{LM^L_{DD^*}}{2\pi}\r)^2}{2\pi}\frac{2}{\pi L}S'\l(-\l(\tfrac{L\gamma^L}{2\pi}\r)^2\r),
	\label{eq:residueL}
\end{align}
where $B^L_\tn{LO}\equiv(\gamma^L)^2/2M^L_{DD^*}$ and
\begin{align}
	S'(x)\equiv\partial_x S(x)=\sum_{\mathbf{n}}\frac{1}{\l(\av{n}^2-x\r)^2}.
	\label{eq:Sprime}
\end{align}
For the coefficient $s_2^L$ we obtain, already inserting the redefinitions of the coupling constants,
\begin{align}
	s_2^L&=c_2(\gamma^L)^2+\frac{g^2}{6(f^L)^2}\l(\frac{M^L_{DD^*}}{2\pi}\r)^2\notag\\
	&\times\l[\l(\gamma-\Lambda\r)^2+(\mu^L)^2\l(-d_2+\log\l(\frac{\mu^{L,\tn{ph}}}{\l|\mu^L\r|}\r)+\frac{1}{2}\r)+\frac{(\mu^L)^2}{4\pi^4}S^\tn{(IV)}\l(-\l(\tfrac{L\gamma^L}{2\pi}\r)^2,\l(\tfrac{L\mu^L}{2\pi}\r)^2\r)\r],
	\label{eq:s2L}
\end{align}
where we used Eq. \eqref{eq:bssol} for the first term in parentheses.

\subsection{Validity range of XEFT in the box}
\label{sec:valrangefinvol}
In the infinite volume, the range of applicability of XEFT is constrained by two demands. On the one hand, we require that pions can be included perturbatively, determining the boundary for large pion masses. On the other hand, treating pions non-relativistically settles the low $m_\pi$ boundary. In summary, we have in the infinite volume $0.98(m_\pi^\tn{ph})^2\lesssim m_\pi^2\lesssim2(m_\pi^\tn{ph})^2$ \cite{Jansen:2013cba}.

\begin{figure}[htbp]
	\begin{center}
		\includegraphics[width=0.7\textwidth]{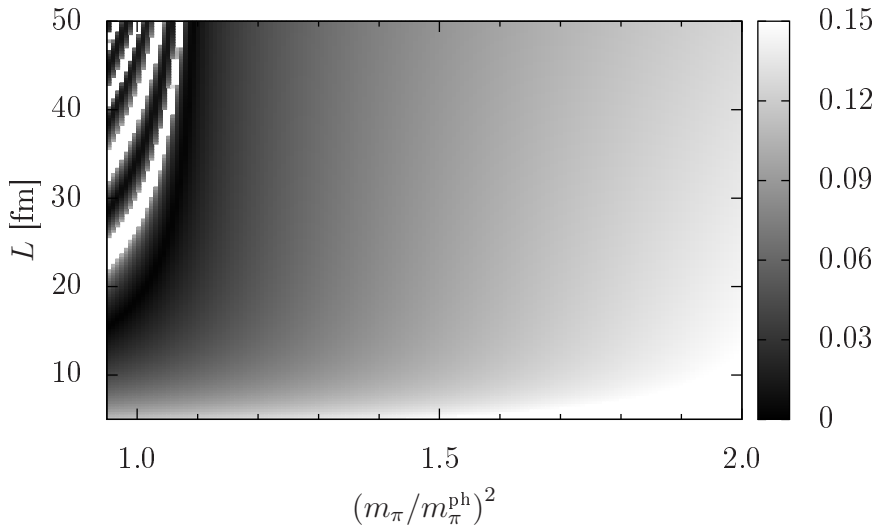}
	\end{center}
	\caption{Density plot for the expansion parameter $\epsilon_\pi$. White regions belong to values of $L$ and $m_\pi$ where the perturbative inclusion of pions breaks down. For pion masses close to the physical value, there are several narrow bands of hyperbolic shape due to on-shell three-body propagators.}
	\label{fig:exppar}
\end{figure}

However, for three particles in the finite volume, singularities occur as soon as three-body propagators can go on-shell, a behavior which has already been investigated e.g. in \cite{Polejaeva:2012ut} and \cite{Briceno:2012rv}. In XEFT, this manifests in the last term of Eq. \eqref{eq:s2L}. For pion masses smaller than the hyperfine splitting, where the $D^{*0}\rightarrow D^0\pi$ decay channel is open, $(\mu^L)^2S^\tn{(IV)}$ possesses singularities for values of $(L\mu^L/2\pi)^2$ being the absolute value of an integer vector squared, greater or equal than one. Since the $D^{*0}\rightarrow D^0\pi$ decay proceeds via a $P$-wave interaction, $s_2^L$ is finite for $(L\mu^L/2\pi)^2=0$. So for certain values of $m_\pi$ and $L$, the perturbative treatment clearly fails. To obtain a region of validity for XEFT in dependence on the volume and the pion mass, we take a look at the quantity
\begin{align}
	\epsilon_\pi\equiv\frac{g^2M^L_{DD^*}}{4\pi (f^L)^2}\l|\frac{1}{\pi L}S\l(\l(\tfrac{L\mu^L}{2\pi}\r)^2\r)+\frac{1}{\pi L}\l(\frac{2\pi}{L\mu^L}\r)^2\r|,
	\label{eq:expfactor}
\end{align}
which explicitly accounts for the singularities of $(\mu^L)^2S^\tn{(IV)}$ for $m_\pi<\Delta$ and approaches the infinite volume XEFT expansion parameter for $m_\pi>\Delta$ and $L\rightarrow\infty$. The second term in Eq. \eqref{eq:expfactor} ensures that $\epsilon_\pi$ is finite for $\mu^L\rightarrow0$. A density plot is shown in Fig. \ref{fig:exppar}.
We restrict our analysis on regions where $\epsilon_\pi<0.15$ such that it is small enough to compensate for unnaturally large NNLO coefficients of similar size as in KSW. For physical pion mass it follows that a perturbative treatment of pions is justified for $5\tn{ fm}\lesssim L\lesssim20\tn{ fm}$. We point out that the NLO parameters $c_2$ and $d_2$ coincide in the finite and infinite volume and, once determined from lattice calculations, one can utilize the infinite volume formulas to extrapolate to $L\rightarrow\infty$.

\subsection{Effective range expansion for large $m_\pi$}
\label{sec:effrangefinvol}
The preceding analysis is valid for all pion masses. Now, we focus on the region where the $D^*$ is stable, i.e. for $m_\pi>\Delta$. Here, we can apply the effective range expansion for the infinite volume amplitude, analytically continue it to negative energies and apply the procedure established in \cite{Beane:2003da}. We introduce the $S$-wave scattering phase shift $\delta_s(p)$, which is related to the infinite volume scattering amplitude by
\begin{align}
	p\cot\delta_s(p)=ip+\frac{2\pi}{M_{DD^*}\mathcal{A}}
	\label{eq:pcotd}
\end{align}
and apply the effective range expansion
\begin{align}
	p\cot\delta_s(p)=-\frac{1}{a_s}+\frac{1}{2}r_sp^2+\cdots.
	\label{eq:effrangeexp}
\end{align}
The quantities $a_s$ and $r_s$ are known as $S$-wave scattering length and $S$-wave effective range, respectively. In the pionless theory, $1/a_s=\gamma$ and $r_s=r_0$. However, including pions leads to corrections of NLO, which can be determined by expanding the inverse infinite volume scattering amplitude in powers of $p$. 
Equating coefficients yields the following expressions for $a_s$ and $r_s$
\begin{subequations}
	\begin{align}
		\frac{1}{a_s}&=\gamma+\frac{g^2}{6f^2}\frac{M_{DD^*}}{2\pi}\l(\l(\gamma-\Lambda\r)^2-\l(\gamma-\l|\mu\r|\r)^2+\l|\mu\r|^2\l(d_2+\frac{1}{2}-\log\l(\frac{\mu^\tn{ph}}{\l|\mu\r|}\r)\r)\r),\label{eq:as}\\
		r_s&=r_0\l[1+\frac{2}{\gamma}\frac{g^2}{6f^2}\frac{M_{DD^*}}{2\pi}\l(\l(\gamma-\Lambda\r)^2-\l(\gamma-\l|\mu\r|\r)^2+\l|\mu\r|^2\l(d_2+\frac{1}{2}-\log\l(\frac{\mu^\tn{ph}}{\l|\mu\r|}\r)\r)\r)\r]\notag\\
		&-\frac{g^2}{6f^2}\frac{M_{DD^*}}{2\pi}2\l(1-\frac{8}{3}\frac{\gamma}{\l|\mu\r|}+2\frac{\gamma^2}{\l|\mu\r|^2}\r).\label{eq:rs}
	\end{align}
\end{subequations}
Let us briefly consider an effective theory in the infinite volume, where pion interactions are not included explicitly but via modified LO and NLO coupling constants with similar renormalization condition as in \eqref{eq:C2renor} with $r_0$ replaced by $r_s$ in Eq. \eqref{eq:rs}. The criteria for a bound state follows from Eq. \eqref{eq:pcotd} and reads, applying Eq. \eqref{eq:effrangeexp} and neglecting higher order shape parameters
\begin{align}
	\frac{1}{a_s}+\frac{1}{2}r_s\gamma_*^2-\gamma_*=0,
	\label{eq:bssoleffrinf}
\end{align}
where $\gamma_*$ is the binding momentum including NLO contributions. Then the binding energy up to NLO is given by
\begin{align}
	E_{X,\tn{NLO}}^\infty&=\frac{1}{2M_{DD^*}}\frac{2}{r_s^2}\l(1-\frac{r_s}{a_s}-\sqrt{1-2\frac{r_s}{a_s}}\r)\notag\\
	&=\frac{1}{2M_{DD^*}a_s^2}\l(1+\frac{r_s}{a_s}+\mathcal{O}\l(\frac{r_s}{a_s}\r)^2\r).
	\label{eq:EXinftyNLO}
\end{align}
Using the same effective theory but now including NLO corrections using the strategy described in Eqs. \eqref{eq:collectedA0} 
through \eqref{eq:BZshift}, we obtain the same result as in the second line of Eq. \eqref{eq:EXinftyNLO} except that no terms of order $r_s^2/a_s^2$ occur. 
Hence, as long as the $S$-wave scattering length is significantly larger than the $S$-wave effective range, both methods deliver consistent results.

Using a pionless effective field theory in the finite volume, the amplitude can be calculated in analogy to Sec. \ref{sec:LOXEFTfinvol} and the result is given by Eq. \eqref{eq:Am1Lreg} with $\gamma$ replaced by $-p\cot\delta_s(p)$. The criteria for a bound state looks similar to \eqref{eq:bssol} and \eqref{eq:bssoleffrinf} (cf. \cite{Beane:2003da})
\begin{align}
	\frac{1}{a_s}+\frac{1}{2}r_s(\gamma_*^L)^2+\frac{1}{\pi L}S\l(-\l(\frac{L\gamma_*^L}{2\pi}\r)^2\r)=0.
	\label{eq:bssoleffr}
\end{align}
Here $\gamma_*^L$ is the finite volume binding momentum including NLO corrections. Eq. \eqref{eq:bssoleffr} approaches Eq. \eqref{eq:bssoleffrinf} in the limit $L\rightarrow\infty$. As for the infinite volume, we expect that the results from the two different methods agree as long as $a_s\gg r_s$.

\section{Results}
\label{sec:results}
In order to determine the finite volume and quark mass dependence of the binding energy, 
we first consider the extrapolations for the pion decay constant, the $D$ meson axial coupling constant and the $D^0$ and $D^{*0}$ meson masses, respectively. 
A superscript (0) denotes the chiral-limit value of a quantity. For the chiral extrapolation of the pion decay constant we use the results from \cite{Gasser:1983yg}
\begin{align}
	f=f^{(0)}\l[1-\frac{1}{4\pi^2{f^{(0)}}^2}m_\pi^2\log\l(\frac{m_\pi}{m_\pi^\tn{ph}}\r)+\frac{\bar{l}_4}{8\pi^2 {f^{(0)}}^2}m_\pi^2\r],
	\label{eq:chiralf}
\end{align}
with the low-energy constant $\bar{l}_4=4.4$ and $f^{(0)}=124\tn{ MeV}$, corresponding to $f^\tn{ph}=132\tn{ MeV}$ \cite{Gasser:1983yg,Colangelo:2001df}. Further, we use the lattice results from \cite{Becirevic:2012pf} for the quark mass dependence of the $D$ meson axial coupling constant
\begin{align}
	g=g^{(0)}\l[1-\frac{1+2{g^{(0)}}^2}{4\pi^2{f^{(0)}}^2}m_\pi^2\log\l(\frac{m_\pi}{\mu_\tn{lat}}\r)+\alpha m_\pi^2\r],
	\label{eq:chiralg}
\end{align}
where the parameters are given as \cite{Becirevic:2012pf}
\begin{align}
	g^{(0)}=0.46,~ ~ ~\alpha=-0.16\tn{ GeV}^{-2},~ ~ ~\mu_\tn{lat}=1\tn{ GeV}.
	\label{eq:gpara}
\end{align}
The $D$ meson axial coupling constant does not receive any corrections in the finite volume. For the pion decay constant we employ the results given in \cite{Gasser:1986vb} obtained from chiral perturbation theory to one loop
\begin{align}
	f^L=f\l[1-\frac{m_\pi}{2\pi f^2}\frac{1}{\pi L}\sum_{\mathbf{n}}^{\av{n}\geq 1}\frac{K_1(\av{n}m_\pi L)}{\av{n}}\r],
	\label{eq:fL}
\end{align}
with $K_1$ being the modified Bessel function of second kind. The chiral and finite volume extrapolations for the $D^0$ and $D^{*0}$ meson masses can be summarized as
\begin{align}
	m^L_D&=m_D=m_D^\tn{ph}+\frac{h_1}{m_D^\tn{ph}}(m_\pi^2-(m_\pi^\tn{ph})^2),\\
	m^L_{D^*}&=m_{D^*}^\tn{ph}+\frac{h_1}{m_{D^*}^\tn{ph}}(m_\pi^2-(m_\pi^\tn{ph})^2)+\Delta_{m_{D^*}}^L,
	\label{eq:Dmesonmasses}
\end{align}
with $h_1=0.42$ \cite{Guo:2009ct} and $\Delta_{m_{D^*}}^L$ given in Eq. \eqref{eq:deltamdstar}.

\begin{figure}[htbp]
	\begin{minipage}{0.49\textwidth}
		\begin{center}
			\includegraphics[width=\textwidth]{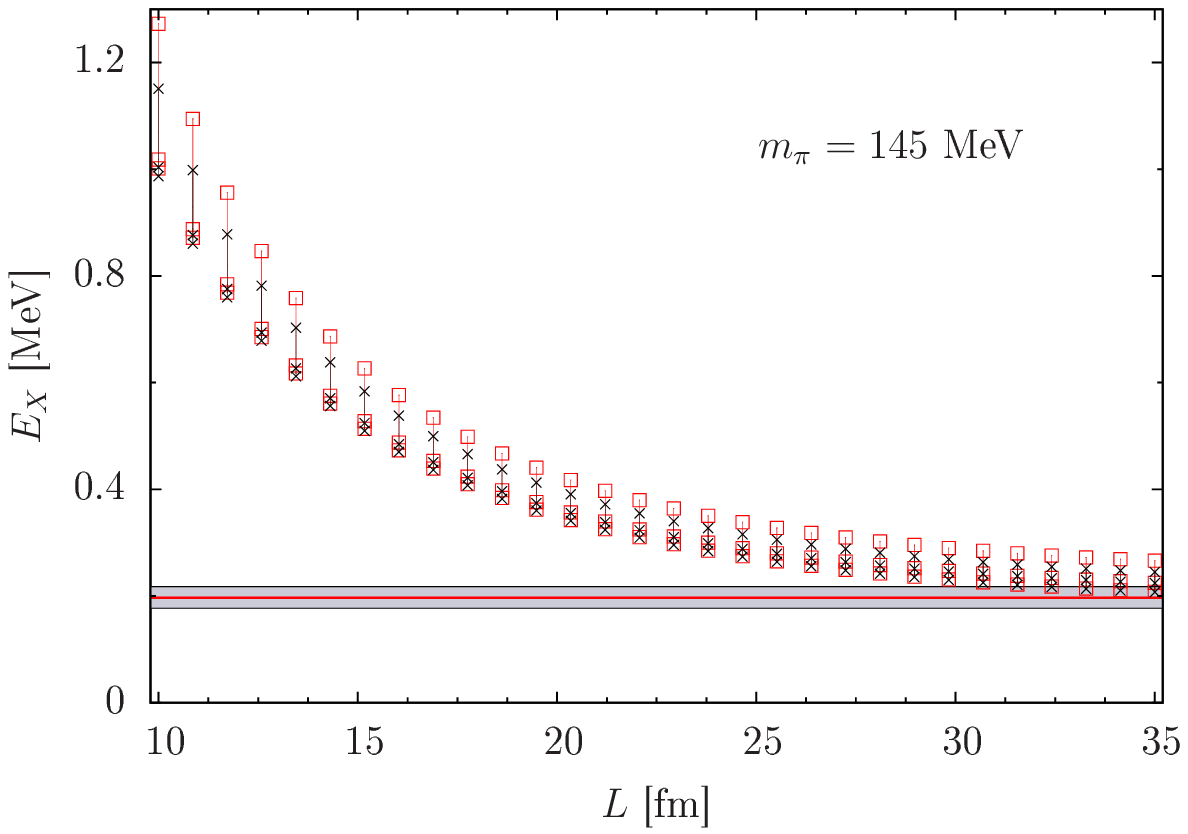}
		\end{center}
	\end{minipage}
	\hfill
	\begin{minipage}{0.49\textwidth}
		\begin{center}
			\includegraphics[width=\textwidth]{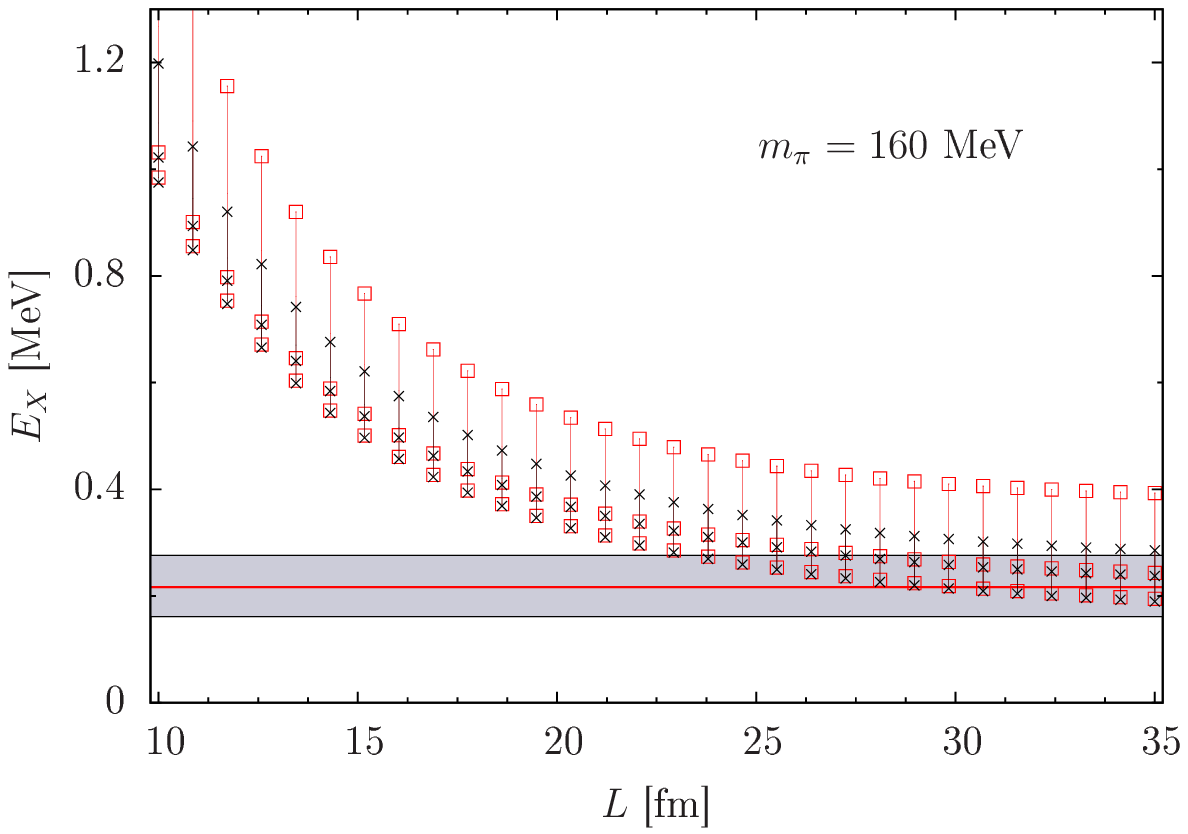}
		\end{center}
	\end{minipage}
	\caption{Comparison of the two methods to obtain the binding energy to NLO described in chapter \ref{sec:BNLO} and \ref{sec:finvol}. We keep the pion mass fixed at values of $m_\pi=145\tn{ MeV}$ (left) and $m_\pi=160\tn{ MeV}$ (right). The binding energy in the infinite volume is represented by solid lines. The band is acquired by varying the NLO parameters $d_2$ and $r_0$ within their natural ranges. The thick, central curve belongs to $d_2=0$ and $r_0=0$. The finite volume results correspond to crosses for the explicit XEFT calculations and to empty squares for the results obtained from an effective range expansion. The central crosses and squares belong to $d_2=0$ and $r_0=0$.}
	\label{fig:BLmp}
\end{figure}

In Fig.~\ref{fig:BLmp}, we plot the dependence of the binding energy on the side length of the box, $L$, to compare the two approaches described in chapter \ref{sec:finvol}. The pion masses are fixed at values of $m_\pi=145\tn{ MeV}$ and $m_\pi=160\tn{ MeV}$, respectively. The infinite volume results are shown by solid lines. The upper bound corresponds to values of the NLO parameters of $d_2=1$ and $r_0=0$ and the lower bound to $d_2=-1$ and $r_0=0$. This results in the maximum band width. In the finite volume, the parameter values for the lower bound are the same but the upper band belongs to $d_2=1$ and $r_0=0.01/\tn{MeV}$ in order to maximize the error band.

Whereas the lower bounds and central values coincide well using the two different strategies and deviations are clearly smaller than the NLO shifts, there is some discrepancy for the upper bounds. This can be understood from the considerations in Sec. \ref{sec:effrangefinvol}. Results are consistent as long as the $S$-wave scattering length is much larger than the $S$-wave effective range. For $d_2=1$ and $r_0=0.01/\tn{MeV}$ however, $a_s$ and $r_s$ are similar in size and the error induced by approximating the root in \eqref{eq:EXinftyNLO} is $\gtrsim10\%$. This is in the order of the NLO corrections and explains the deviation for the upper bounds in Fig. \ref{fig:BLmp}. We point out that effective range and scattering length are of comparable magnitude for a very limited range of the NLO parameters only.

\begin{figure}[htbp]
	\begin{center}
		\includegraphics[width=0.7\textwidth]{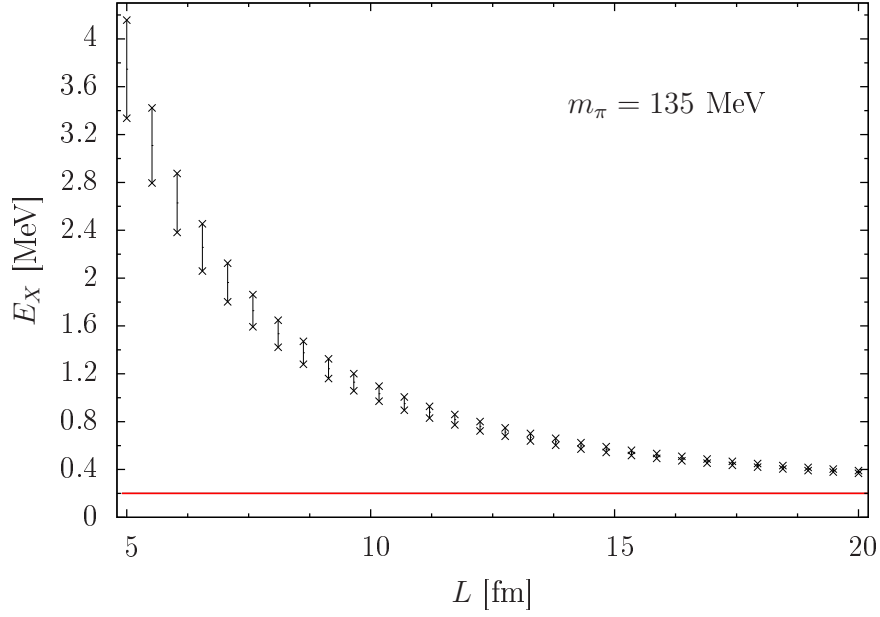}
	\end{center}
	\caption{Volume dependence of the binding energy for physical quark masses. We use the same notation as in Fig. \ref{fig:BLmp}. Note that the effective range expansion breaks down for pion masses below $142\tn{ MeV}$ and hence effective range results are not included. Furthermore, we renormalized the binding energy in the infinite volume to $0.2\tn{ MeV}$ and thus the infinite volume result is represented by a single line.}
	\label{fig:BLmp135}
\end{figure}

So far we looked at pion masses above the hyperfine splitting of the $D$-mesons. We now consider the region where the $D^{*0}$ can decay into $D^0\pi$. The binding energy in a finite volume for physical pion mass is depicted in Fig. \ref{fig:BLmp135}. We plot for box lengths between $5$ and $20\tn{ fm}$ where the expansion parameter in Eq. \eqref{eq:expfactor} is clearly smaller than $0.15$ as can be read off from Fig. \ref{fig:exppar}.

The result for $d_2=0$ and $r_0=0$ is not shown as it almost coincides with the lower bound. This can be understood by noting that the only difference of the central values and the lower bound is the value of $d_2$. Since $\mu$ is rather insensitive to effects of the finite volume, the NLO contact interaction with vertex $-iD_2\mu^2$ barely differs in a finite box. The renormalization to $E_X=0.2\tn{ MeV}$ at physical pion mass then explains the similarity of the outcome for $d_2=0$ and $d_2=-1$. The renormalization condition further explains why there is no error band for the binding energy in the infinite volume for $m_\pi=135\tn{ MeV}$. The contribution of the NLO contact interaction with coupling constant $C_2$ on the other hand is proportional to $(\gamma^L)^2$ and since finite volume corrections to $\gamma$ are significantly greater than those to $\mu$, 
the error band for physical pion mass is predominantly determined by $r_0$.

\begin{figure}[htbp]
	\begin{center}
		\includegraphics[width=0.7\textwidth]{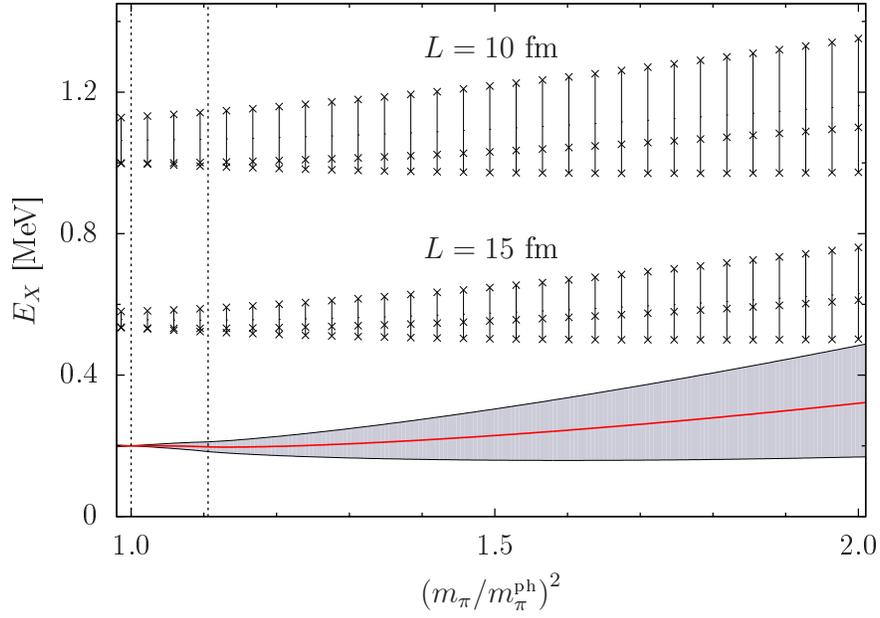}
	\end{center}
	\caption{Quark mass dependence of the binding energy for various box lengths. From top to bottom, box lengths of $L=10\tn{ fm}$, $L=15\tn{ fm}$ and $L=\infty$ are accounted for. We use the same notation as in Fig. \ref{fig:BLmp}. Again, no effective range results are included since these are not valid over the whole range of the quark masses.}
	\label{fig:BLL}
\end{figure}
The binding energy is, as expected, increasing for decreasing box size and approaching the infinite volume value for large volumes. However, even at $L=20\tn{ fm}$ finite volume contributions are still above $50\%$. The $X$ is significantly deeper bound for small box lengths and finite volume corrections yield the dominating contribution to the binding energy. Besides the demand that pions can be included perturbatively, it is required that the binding momentum does not exceed the scales integrated out, which are at 
the order of the pion mass. For a volume with $L\gtrsim5\tn{ fm}$, $E_X\lesssim4\tn{ MeV}$ corresponding to binding momentum $\lesssim90\tn{ MeV}$, we expect 
that XEFT properly describes the dynamics of the $X$.

The chiral extrapolations for fixed box size of $L=10\tn{ fm}$, $L=15\tn{ fm}$ and $L=\infty$ are shown in Fig. \ref{fig:BLL}. The infinite volume results are again depicted by solid lines. The NLO parameters for the bounds coincide with the ones for Fig. \ref{fig:BLmp}. As in the infinite volume, the binding energy in a finite box shows only a moderate 
sensitivity to the light quark masses. The central values of the finite volume belong to $d_2=0$ and $r_0=0$ and approach the lower bound for physical pion mass for reasons explained above.

\section{Conclusion and Outlook}
\label{sec:conout}

In this work, we examined the $X(3872)$ in a finite volume using XEFT to NLO. We combined our results with chiral extrapolations, which we derived in an earlier paper \cite{Jansen:2013cba}. A feature of XEFT is that NLO interactions can be included perturbatively as long as the expansion parameter for the inclusion of pions is sufficiently small. Based on rather conservative assumptions, we estimated domains for the light quark masses and, in the finite volume, for the box size, where XEFT is expected to remain valid.

In these domains, we gave explicit expressions for finite volume corrections to the binding energy. On the one hand, we utilized a method, used in the infinite volume as well, which can be applied for all considered values of the quark masses, even those for which the $D^{*0}$ can decay. On the other hand, for stable $D^{*0}$, 
we further employed an approach implementing the effective range expansion, which served as a consistency check.

Moreover, we showed that the finite volume shift to the binding energy is fully determined by infinite volume parameters and no additional input is needed. By implication this means that the two undetermined parameters of XEFT to NLO, denoted by $r_0$ and $d_2$, can be determined from lattice calculations. Here we estimated natural ranges and varied them within these to determine the error bands. Although there are certain values of $r_0$ and $d_2$, where the results of the two strategies mentioned above deviate and the error bars are possibly underestimated, over most of the natural ranges for the NLO parameters both methods yield consistent results.

For all examined values of box sizes and light quark masses, we found that finite volume corrections play a crucial role and yield shifts being at least in the order of the physical binding energy. Furthermore, over the whole natural ranges of the NLO parameters $r_0$ and $d_2$, the $\bar{D}^0D^{*0}$ system is bound. From these findings, we conclude that the $X$ should be observable on the lattice and already 
at box lengths $L\sim 20$ fm
is expected to be a factor two more deeply bound than experimentally measured.

Our analysis could be used in order to extrapolate results of lattice simulations to physical quark masses and infinite volumes.
In the first full lattice QCD study of the $X(3872)$ in 2013,
Prelovsek and Leskovec \cite{Prelovsek:2013cra} identified a state $(11\pm7)\tn{ MeV}$ below the $\bar{D}^0D^{*0}$ threshold with the $X$
for squared pion masses about four times the physical value and a spatial box size of $2\tn{ fm}$. This pion mass is clearly beyond the range of applicability of XEFT.
Ignoring this problem and extrapolating the physical binding energy to this volume and pion mass using our results, we find the value
$E_X(m_\pi=270\tn{ MeV}, L=2\tn{ fm})\approx 20\tn{ MeV}$ which is roughly consistent with the lattice result.
A newer lattice study from 2015 \cite{Padmanath:2015era} arrives at a similar result as Prelovsek and Leskovec \cite{Prelovsek:2013cra}
as does the preliminary outcome from Lee et al. \cite{Lee:2014uta}.

Among the outstanding challenges are the systematic incorporation of discretization effects, an analysis of the impact of unphysical charm quark masses
and an improved understanding of the effect of $c\bar{c}$ operators \cite{Mohler:2015bet}.
Finally, an analysis of coupled channel effects, in particular the determination of the ${}^3S_1-{}^3D_1$ mixing angle,
e.g.~by following the strategy described for nucleons in \cite{Briceno:2013bda}, remains for future work.
An analysis of the finite volume corrections in the Galilean-invariant version of XEFT~\cite{Braaten:2015tga} would also be
interesting.

\begin{acknowledgments}
We thank A. Rusetsky for helpful discussions.
This research was supported in part by the 
Helmholtz Association under contract HA216/EMMI,
by the National Natural Science Foundation of China under Grants No.~11261130311 (CRC110 by DFG and NSFC), 
and No.~11475188.

\end{acknowledgments}

\numberwithin{equation}{section}
\appendix
\section{Next-to-leading order scattering diagrams}
\label{sec:NLOsa}
The NLO off-shell scattering diagrams depicted in Fig. \ref{fig:A0} read
\begin{subequations}
\begin{align}
	i\mathcal{A}_0^\tn{(I)}=&-i\frac{C_2}{2}(\avb{\ell}^2+\avb{\ell'}^2)-2i\frac{C_2}{2}(\tfrac{1}{2}(\avb{\ell}^2+\avb{\ell'}^2)-\eta^2)(-\eta+\Lambda)\frac{M_{DD^*}}{2\pi}\mathcal{A}_{-1}\notag\\
	&-i\frac{C_2}{2}(-2\eta^2)(-\eta+\Lambda)^2\l(\frac{M_{DD^*}}{2\pi}\r)^2(\mathcal{A}_{-1})^2,\label{eq:a0Ires}\\
	i\mathcal{A}_0^\tn{(II)}=&\frac{ig^2}{6f^2}\l(1+\frac{\mu^2}{4\avb{\ell}\avb{\ell'}}\log\l(1-\frac{4\avb{\ell}\avb{\ell'}}{\mu^2-\l|\boldsymbol{\ell}-\boldsymbol{\ell'}\r|^2}\r)\r),\label{eq:a0IIres}\\
	i\mathcal{A}_0^\tn{(III)}=&\frac{ig^2}{3f^2}\l(\l(-\eta+\Lambda\r)+\frac{i\mu^2}{2}\l(\frac{1}{2\avb{\ell}}\log\l(1+\frac{2\avb{\ell}}{i\eta+\mu-\avb{\ell}}\r)+\avb{\ell}\longleftrightarrow \avb{\ell'}\r)\r)\frac{M_{DD^*}}{2\pi}\mathcal{A}_{-1},\label{eq:a0IIIres}\\
	i\mathcal{A}_0^\tn{(IV)}=&\frac{ig^2}{6f^2}\l(\l(-\eta+\Lambda\r)^2+\mu^2\l(\log\l(\frac{\Lambda}{2\eta-i\mu}\r)+\frac{1}{2}+R\r)\r)\l(\frac{M_{DD^*}}{2\pi}\r)^2(\mathcal{A}_{-1})^2,\label{eq:a0IVres}\\
	i\mathcal{A}_0^\tn{(V)}=&\frac{-iD_2\mu^2}{(C_0)^2}(\mathcal{A}_{-1})^2,\label{eq:a0Vres}
\end{align}
\end{subequations}
where $\boldsymbol\ell$ and $\boldsymbol\ell'$ are off-shell momenta while $\eta\equiv\sqrt{-2M_{DD^*}(E+\Sigma^\tn{OS})-i\epsilon}$ depends on the energy.
Moreover,
$R\equiv\tfrac{1}{2}(-\gamma_E+\log\l(\pi\r))$ is a renormalization constant, which we absorb in the NLO coupling constant $D_2$ (cf. Eq. \eqref{eq:D2renor}).

\section{Next-to-next-to-leading order cancellations}
\label{sec:NNLOcancel}
\begin{figure}[htbp]
	\begin{center}
		\includegraphics[width=0.8\textwidth]{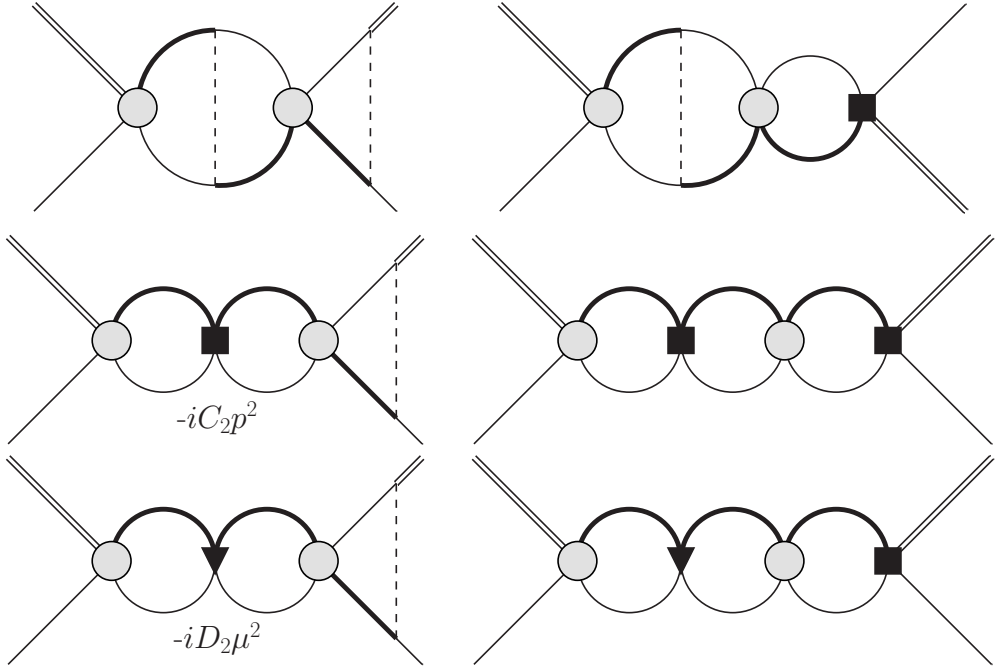}
	\end{center}
	\caption{Scattering amplitudes shifting the coefficient $s_2$ in Eq. \eqref{eq:collectedA0} at NNLO and causing a cancellation in Eq. \eqref{eq:Bshift}.}
	\label{fig:s2nnlo}
\end{figure}

Starting from Eqs. \eqref{eq:fullamplitude} and \eqref{eq:nonpertamplitude}, we first compute the shift of the residue up to NLO by equating the terms linear in the LO amplitude and obtain Eq. \eqref{eq:Zshift}. Plugging the shifted residue in Eq. \eqref{eq:nonpertamplitude} yields for the shift of the binding energy
\begin{align}
	\Delta B^\tn{NLO}=-\frac{Z_{-1}}{1+s_1}s_2.
	\label{eq:deltaBdirect}
\end{align}
The dimensionless coefficient $s_1$ derives from the NLO amplitudes and is thus expected to be much smaller than $1$. Hence, we expand the overall factor $1/(1+s_1)$ in Eq. \eqref{eq:deltaBdirect} as a geometrical series and obtain
\begin{align}
	\Delta B^\tn{NLO}=-Z_{-1}\l(s_2-s_1s_2+\mathcal{O}(s_1^2s_2)\r).
	\label{eq:expansiondeltaB}
\end{align}
Taking into account that $s_2$ is of NLO, too, we anticipate that terms proportional to $s_1s_2$ are actually of NNLO. In fact, at NNLO there are six diagrams proportional to the LO amplitude squared, depicted in Fig. \ref{fig:s2nnlo}, leading to an NNLO shift for the coefficient $s_2$, which exactly cancels out the second term in Eq. \eqref{eq:expansiondeltaB}. This can be seen utilizing that amplitudes separate at resummed LO vertices. It is because of this cancellation that we omit the factor $1/(1+s_1)$ in Eq. \eqref{eq:Bshift}.

\section{Calculation of the one-pion exchange amplitude $\mathcal{A}_0^\tn{(IV)}$}
\label{sec:A0IV}
The OPE amplitude $\mathcal{A}_0^\tn{(IV)}$ is depicted in Fig. \ref{fig:A0}. We begin with the unregularized expression in the infinite volume
\begin{align}
	&i\hat{\mathcal{A}}^\tn{(IV)}_{0\,ij}=\notag\\
	&-i\mathcal{A}_{-1}^2\frac{g^2}{2f^2}\frac{1}{2m_\pi}\int\frac{dk_0}{2\pi i}\int\frac{dk'_0}{2\pi i}\int\frac{d^3\mathbf{k}}{(2\pi)^3}\int\frac{d^3\mathbf{k'}}{(2\pi)^3}\frac{1}{E+k_0-\av{k}^2/2m_{D^*}+i\epsilon}~\frac{1}{k_0+\av{k}^2/2m_D-i\epsilon}\notag\\
	&\cdot\frac{\boldsymbol{\varepsilon}_i\cdot(\mathbf{k}+\mathbf{k'})~\boldsymbol{\varepsilon}^*_j\cdot(\mathbf{k}+\mathbf{k'})}{E+k_0+k'_0-\l|\mathbf{k}+\mathbf{k'}\r|^2/2m_\pi+\delta+i\epsilon}~\frac{1}{E+k'_0-\av{k'}^2/2m_{D^*}+i\epsilon}~\frac{1}{k'_0+\av{k'}^2/2m_D-i\epsilon}.
	\label{eq:A0IVinf}
\end{align}
To transition into the finite volume we replace the spatial integration by sums over the allowed lattice momenta
\begin{align}
	\int\frac{d^3\mathbf{k}}{(2\pi)^3}\xrightarrow{V\rightarrow L^3}\frac{1}{L^3}\sum_{\mathbf{k}=\frac{2\pi}{L}\mathbf{n}}.
	\label{eq:transitiontoL}
\end{align}
At the same time we keep the contour integration over the time component since lattice simulations are usually performed with significantly larger time than spatial interval. We acquire
\begin{align}
	i\mathcal{A}_{0\tn{(IV)}ij}^L=i(\mathcal{A}_{-1}^L)^2\frac{g^2}{2(f^L)^2}\frac{1}{L^3}\sum_{\mathbf{k}=\frac{2\pi}{L}\mathbf{n}}\frac{1}{L^3}\sum_{\mathbf{k'}=\frac{2\pi}{L}\mathbf{n'}}&\frac{1}{E-\av{k}^2/2M^L_{DD^*}}~\frac{1}{E-\av{k'}^2/2M^L_{DD^*}}\label{eq:A0IVL}\\
	\cdot&\frac{\boldsymbol{\varepsilon}_i\cdot(\mathbf{k}+\mathbf{k'})~\boldsymbol{\varepsilon}^*_j\cdot(\mathbf{k}+\mathbf{k'})}{\l|\mathbf{k}+\mathbf{k'}\r|^2-2m_\pi\delta^L-2m_\pi E-\frac{m_\pi}{m_D}(\av{k}^2+\av{k'}^2)}\notag.
\end{align}
As a next step, we evaluate at an energy $E=p^2/2M^L_{DD^*}$, neglect terms proportional to $m_\pi/m_D$ and $\delta^L/m_\pi$, respectively, and use a tensor decomposition to replace
\begin{align}
	\boldsymbol{\varepsilon}_i\cdot(\mathbf{n}+\mathbf{n'})~\boldsymbol{\varepsilon}_j\cdot(\mathbf{n}+\mathbf{n'})\rightarrow\frac{\delta_{ij}}{3}\l|\mathbf{n}+\mathbf{n'}\r|^2.
	\label{eq:replacedelta}
\end{align}
We get for the scalar amplitude
\begin{align}
	i\mathcal{A}_{0\tn{(IV)}}^L\approx i(\mathcal{A}_{-1}^L)^2\frac{g^2}{6(f^L)^2}\l(\frac{M^L_{DD^*}}{2\pi}\r)^2\frac{1}{(\pi L)^2}\sum_{\mathbf{n},\mathbf{n'}}\frac{1}{\av{n}^2-\l(\frac{Lp}{2\pi}\r)^2}~\frac{1}{\av{n'}^2-\l(\frac{Lp}{2\pi}\r)^2}\frac{\l|\mathbf{n}+\mathbf{n'}\r|^2}{\l|\mathbf{n}+\mathbf{n'}\r|^2-\l(\frac{L\mu^L}{2\pi}\r)^2}.
	\label{eq:scalarA0IVL}
\end{align}
To renormalize, we introduce a cut-off $\lambda_n$ for the sum, add and subtract the infinite volume loop integrals evaluated at zero energy, one regularized using a cut-off and the other one using PDS and take the limit $\lambda_n\rightarrow\infty$

\begin{align}
	i\mathcal{A}_{0\tn{(IV)}}^L=&i(\mathcal{A}_{-1}^L)^2\frac{g^2}{6(f^L)^2}\l(\frac{M^L_{DD^*}}{2\pi}\r)^2\biggl[\l(\frac{1}{\pi L}S\l(\l(\tfrac{Lp}{2\pi}\r)^2\r)+\Lambda\r)^2\notag\\
	+&(\mu^L)^2\biggl(\frac{1}{(2\pi^2)^2}S^\tn{(IV)}\l(\l(\tfrac{Lp}{2\pi}\r)^2,\l(\tfrac{L\mu^L}{2\pi}\r)^2\r)+\log\l(\frac{\Lambda}{\l|\mu^L\r|}\r)+\frac{1}{2}+R\biggr)\biggr],
	\label{eq:ALIVreg}
\end{align}
where $S(x)$ is defined in Eq. \eqref{eq:Sx} and $S_\tn{(IV)}(x,y)$ is given as
\begin{align}
	S^\tn{(IV)}(x,y)\equiv\lim_{\lambda_n\rightarrow\infty}\Biggl[&\sum_{\mathbf{n},\mathbf{n'}}^{\av{n},\av{n'}<\lambda_n}\frac{1}{\av{n}^2-x}~\frac{1}{\av{n'}^2-x}~\frac{1}{\l|\mathbf{n}+\mathbf{n'}\r|^2-y}\notag\\
	&-2\pi^4\l(\log\l(\frac{\lambda_n^2}{\l|y\r|}\r)-1+\mathcal{O}\l(\log\l(\frac{\lambda_n^2}{\l|y\r|}\r)\cdot\frac{\l|y\r|}{\lambda_n^2}\r)\r)\Biggr].
	\label{eq:SIV}
\end{align}
For the cut-off regularized integral, we expanded in $\mu/\lambda$
\begin{align}
	&\int d^3\mathbf{k}\int d^3\mathbf{k'}~\frac{\theta(\lambda-\av{k})}{q^2}\frac{\theta(\lambda-\av{k'})}{k^2}\frac{1}{\l|\mathbf{k}+\mathbf{k'}\r|^2-\mu^2-i\epsilon}\notag\\
	=&\log\l(\frac{\lambda^2}{\mu^2}\r)\l(I_0^{(1)}+I_1^{(1)}\frac{\mu}{\lambda}+I_2^{(1)}\frac{\mu^2}{\lambda^2}+\mathcal{O}\l(\frac{\mu^3}{\lambda^3}\r)\r)+I_0^{(2)}+I_1^{(2)}\frac{\mu}{\lambda}+I_2^{(2)}\frac{\mu^2}{\lambda^2}+\mathcal{O}\l(\frac{\mu^3}{\lambda^3}\r)
	\label{eq:SIVcutoff}
\end{align}
and found for the coefficients
\begin{alignat}{2}
	&I_0^{(1)}=2\pi^4, ~ ~ ~ &&I_0^{(2)}=2\pi^4\l(-1+i\pi\r),\notag\\
	&I_1^{(1)}=0, ~ ~ ~ &&I_1^{(2)}=-8i\pi^3,\notag\\
	&I_2^{(1)}=-2\pi^2, ~ ~ ~ &&I_2^{(2)}=-2\pi^2\l(1+i\pi+\log(4)\r).
	\label{eq:expansioncoeffSIV}
\end{alignat}

\bibliography{ref}


\end{document}